\documentstyle[11pt,aaspp4,flushrt]{article}

\begin{document}

\title{Correlated intense X-ray and TeV activity \\
of Mrk~501 in 1998 June} 

\author{R.M.~Sambruna\altaffilmark{1}, 
F.A.~ Aharonian\altaffilmark{2}, 
H.~Krawczynski\altaffilmark{2},
\\[3ex]} 
\author{
A.G.~Akhperjanian\altaffilmark{8},
J.A.~Barrio\altaffilmark{3,4},
K.~Bernl\"ohr\altaffilmark{2,5},
H.~Bojahr\altaffilmark{7},
I.~Calle\altaffilmark{4},
J.L.~Contreras\altaffilmark{4},
J.~Cortina\altaffilmark{4},
S.~Denninghoff\altaffilmark{3},
V.~Fonseca\altaffilmark{4},
J.C.~Gonzalez\altaffilmark{4},
N.~G\"otting\altaffilmark{5},
G.~Heinzelmann\altaffilmark{5},
M.~Hemberger\altaffilmark{2},
G.~Hermann\altaffilmark{2},
A.~Heusler\altaffilmark{2},
W.~Hofmann\altaffilmark{2},
D.~Horns\altaffilmark{5},
A.~Ibarra\altaffilmark{4},
R.~Kankanyan\altaffilmark{2,8},
M.~Kestel\altaffilmark{3},
J.~Kettler\altaffilmark{2},
C.~K\"ohler\altaffilmark{2},
A.~Kohnle\altaffilmark{2},
A.~Konopelko\altaffilmark{2},
H.~Kornmeyer\altaffilmark{3},
D.~Kranich\altaffilmark{3},
H.~Lampeitl\altaffilmark{2},
A.~Lindner\altaffilmark{5},
E.~Lorenz\altaffilmark{3},
N.~Magnussen\altaffilmark{7},
O.~Mang\altaffilmark{6},
H.~Meyer\altaffilmark{7},
R.~Mirzoyan\altaffilmark{3},
A.~Moralejo\altaffilmark{4},
L.~Padilla\altaffilmark{4},
M.~Panter\altaffilmark{2},
R.~Plaga\altaffilmark{3},
A.~Plyasheshnikov\altaffilmark{2},
J.~Prahl\altaffilmark{5},
G.~P\"uhlhofer\altaffilmark{2},
G.~Rauterberg\altaffilmark{6},
A.~R\"ohring\altaffilmark{5},
V.~Sahakian\altaffilmark{8},
M.~Samorski\altaffilmark{6},
M.~Schilling\altaffilmark{6},
D.~Schmele\altaffilmark{5},
F.~Schr\"oder\altaffilmark{7},
W.~Stamm\altaffilmark{6},
M.~Tluczykont\altaffilmark{5},
H.J.~V\"olk\altaffilmark{2},
B.~Wiebel-Sooth\altaffilmark{7},
C.~Wiedner\altaffilmark{2},
M.~Willmer\altaffilmark{6},
W.~Wittek\altaffilmark{3}, The HEGRA collaboration,\\[3ex]}
\author{L.~Chou\altaffilmark{1}, P.S.~Coppi\altaffilmark{9}, 
R.~Rothschild\altaffilmark{10}, C.M.~Urry\altaffilmark{11}}
\altaffiltext{1}{Pennsylvania State University,
Department of Astronomy and Astrophysics, 
525 Davey Lab, University Park, PA 16802}
\altaffiltext{2}{Max Planck Institut f\"ur Kernphysik,
Postfach 103980, D-69029 Heidelberg, Germany}
\altaffiltext{3}{Max Planck Institut f\"ur Physik, F\"ohringer Ring
6, D-80805 M\"unchen, Germany}
\altaffiltext{4}{Universidad Complutense, Facultad de Ciencias
F\'{\i}sicas, Ciudad Universitaria, E-28040 Madrid, Spain }
\altaffiltext{5}{Universit\"at Hamburg, II. Institut f\"ur
Experimentalphysik, Luruper Chaussee 149,
D-22761 Hamburg, Germany}
\altaffiltext{6}{Universit\"at Kiel, Institut f\"ur Experimentelle und 
Angewandte Physik,
Leibnizstra{\ss}e 15-19, D-24118 Kiel, Germany}
\altaffiltext{7}{Universit\"at Wuppertal, Fachbereich Physik,
Gau{\ss}str.20, D-42097 Wuppertal, Germany}
\altaffiltext{8}{Yerevan Physics Institute, Alikhanian Br. 2, 375036 Yerevan, 
Armenia}
\altaffiltext{9}{Yale University, New Haven, CT 06520-8101}
\altaffiltext{10}{Center for Astrophysics and Space Sciences, 
University of California at San Diego, La Jolla, CA 92093}
\altaffiltext{11}{STScI, 3700 San Martin Dr., Baltimore, MD 21218}

\begin{abstract}
We present exactly simultaneous X-ray and TeV monitoring with {\it
RXTE} and HEGRA of the TeV blazar Mrk 501 during 15 days in 1998 June.
After an initial period of very low flux at both wavelengths, the
source underwent a remarkable flare in the TeV and X-ray energy bands, 
lasting for about six days and with a larger amplitude at 
TeV energies than in the X-ray band.
At the peak of the TeV flare, rapid TeV flux variability on sub-hour
timescales is found. 
Large spectral variations are observed at X-rays, with the 3--20~keV
photon index of a pure power law continuum flattening from
$\Gamma=2.3$ to $\Gamma=1.8$ on a timescale of 2--3 days. This implies
that during the maximum of the TeV activity, the synchrotron peak
shifted to energies $\gtrsim 50$ keV, a behavior similar to that
observed during the longer-lasting, more intense flare in 1997
April. The TeV spectrum during the flare is described by a 
power law with photon index $\Gamma=1.9$ and an exponential cutoff at
$\sim$ 4~TeV; an indication for spectral softening during the flare
decay is observed in the TeV hardness ratios. Our results generally
support a scenario where the TeV photons are emitted via inverse
Compton scattering of ambient seed photons by the same electron
population responsible for the synchrotron X-rays. The simultaneous
spectral energy distributions (SEDs) can be fit with a one-zone
synchrotron-self Compton model assuming a substantial increase of the
magnetic field and the electron energy by a factor of 3 and 10,
respectively.
\end{abstract}

\noindent {\underline{\em Subject Headings:}} Galaxies:jets --
X-rays:galaxies -- Radiation mechanisms:non-thermal -- BL Lacertae
objects:Mrk 501 --- Gamma rays:observations.

\section{Introduction} 

BL Lacertae objects (BL Lacs) are radio-loud AGN dominated by
non-thermal continuum emission from radio up to $\gamma$-rays (MeV to
TeV energies) from a relativistic jet oriented at small angles to the
observer (e.g., Urry \& Padovani 1995). While the radio through
UV/X-ray continuum is almost certainly due to synchrotron emission
from relativistic electrons in the jet (Ulrich, Maraschi, \& Urry 1997
and references therein), the origin of the luminous $\gamma$-ray
radiation from BL Lacs is still uncertain. Possibilities include
inverse Compton scattering of ambient photons off the jet electrons
(Maraschi et al. 1992; Sikora, Begelman, \& Rees 1994; Dermer et
al. 1996), or hadronic processes (e.g.\ Dar \& Laor 1997; Mannheim
1993). 

A breakthrough was provided by the discovery of TeV emission from a
handful of such sources, all characterized by a synchrotron peak at
higher energies (High-energy peaked BL Lacs, or HBLs). One of these is
Mrk 501 ($z$=0.034). This source came into much attention after it
exhibited a prolonged period of intense TeV activity in 1997 (Catanese
et al. 1997; Hayashida et al.\ 1998; Quinn et al.\ 1999; Aharonian et
al. 1997,1999a-c; Djannati-Atai et al. 1999), accompanied by
correlated X-ray emission on timescales of days. Interestingly, this
exceptional TeV activity was accompanied by unusually hard X-ray
emission up to $\gtrsim$ 100 keV (Pian et al. 1998a; Catanese et
al. 1997; Lamer \& Wagner 1999; Krawczynski et al.\ 1999),
unprecedented in this or any other BL Lac. The hard X-ray spectrum
implied a shift toward higher energies of the synchrotron peak,
usually located at UV/soft X-rays (e.g., Sambruna, Maraschi, \& Urry
1996; Kataoka et al. 1999), by more than three decades, persistent
over a timescale of $\sim$ 10 days (Pian et al. 1998a). Further
observations with {\it BeppoSAX} in April-May 1998 and in May 1999
during periods of TeV lower flux showed that the synchrotron peak had
decreased to $\sim$ 20 and 0.5 keV, respectively (Pian et al. 1998b,
1999). These secular variations of the synchrotron peak suggest a
powerful mechanism of particle energization, operating over timescales
of years.
 
Because of its bright TeV emission and unusual X-ray spectral
properties, we selected Mrk 501 for an intensive monitoring in 1998
June using HEGRA and the {\it Rossi X-ray Timing Explorer} ({\it
RXTE}), with a sampling designed to probe correlated variability at
the two wavelengths on timescales of one day or shorter. Here we
report the first results of the campaign, which is characterized by
the detection of a strong flare at both TeV and X-ray energies after a
period of very low activity. The structure of this paper is as
follows.  We describe the sampling and the observations in \S~2, the
X-ray and TeV light curves in \S~3.1, and the TeV and X-ray spectra in
\S\S~3.2--3.3.  Implications of the data are discussed in \S~4.

\section{Sampling and Data Analysis}  

The {\it RXTE} observations of Mrk 501 started June 14 and ended June
28, with a sampling of once per day. The exposure time, typically 2--7
ks during the first week of observations (as allowed by visibility),
decreased to 0.5--1 ks during the latest period of the campaign, due
to reduced visibility constraints. The total exposure in 1998 June was
45,184 s. The remaining 134 ks of the total allocated exposure were 
re-scheduled in 1998 July and August; these data will be presented in
a future publication, together with simultaneous observations at
longer wavelengths (Sambruna et al. 2000).  The HEGRA observations
started one day earlier and ended three days later than {\it RXTE},
with typical integration times of 1.5--2 hours per night, covering
100\% of the {\it RXTE} exposure.

\subsection{X-ray observations}

The {\it RXTE} data were collected in the 2--60 keV band with the
Proportional Counter Array (PCA; Jahoda et al. 1996) and in the
15--250 keV band with the High-Energy X-ray Timing Experiment (HEXTE;
Rothschild et al. 1998). For the best signal-to-noise ratio,
Standard-2 mode PCA data gathered with the top layer of the operating
PCUs 0, 1, and 2 were analyzed.  The data were extracted using the
script \verb+REX+ which adopts standard screening criteria; the net
exposure after screening in each Good Time Interval ranges from 0.2 to
6 ks (Table 2; see below). The background was evaluated using models
and calibration files provided by the {\it RXTE} GOF for a ``faint''
source (less than 40 c/s/PCU), using \verb+pcabackest+ v.2.1b.  Light
curves were extracted in various energy ranges to study the
energy-dependence of the flux variability; for simplicity, only the
light curves in 2--4 keV and 10--20 keV (at the two extrema of the
total energy range of the PCA) will be shown here.

The HEXTE data were extracted from both clusters for the same time
periods as the PCA. Due to the weak nature of the hard X-ray flux, the
data were combined into pre-flare (MJD 50980--988) and flare (MJD
50989--993) time intervals. In addition, the flare interval was
further subdivided into the rising portion (MJD 50989--990) and the
rest of the flare containing the peak intensity. The source signal is
detected to about 50 keV, and we present results from these average
spectra only.


Response matrices for the PCA data were created with \verb+PCARMF+
v.3.5.  Spectral analysis of the PCA and HEXTE data was performed
within \verb+XSPEC+ v.10.0, using the latest released versions of the
spectral response files. The fits were performed in the energy ranges
3--20 keV and 20--250 keV, where the calibrations are best known.  The
quoted uncertainties on the spectral parameters are 90\% confidence
for one parameter of interest ($\Delta\chi^2$=2.7).

\subsection{TeV observations}

The HEGRA Cherenkov telescope system (Daum et al. 1997; Konopelko et
al. 1999) is located on the Roque de los Muchachos on the Canary
Island of La Palma (lat.\ 28.8$^\circ$ N, long.\ 17.9$^\circ$ W, 2200
m a.s.l.).  The Mrk~501 observations described in this paper were
taken from June 14th, 1998 to July 3rd, 1998 and comprise 49~hours of
best quality data.  The analysis tools, the procedure of data cleaning
and fine tuning of the Monte Carlo simulations, as well as the
estimate of the systematic errors on the differential $\gamma$-ray
energy spectra, were discussed in detail by Aharonian et al. (1999a,b).

The analysis uses the standard ``loose'' $\gamma$/hadron separation
cuts which minimize systematic errors on flux and spectral estimates
rather than yielding the optimal signal-to-noise ratio.  A software
requirement of two IACTs within 200~m from the shower axis, each with
more than 40 photoelectrons per image and a ``distance'' parameter of
smaller than 1.7$^\circ$ was used.  Additionally, only events with a
minimal stereo angle larger than 20$^\circ$ were admitted to the
analysis.  Integral fluxes above a certain energy threshold were
obtained by integrating the differential energy spectra above the
threshold energy, rather than by simply scaling detection rates. By
this means integral fluxes were computed without assuming a certain
source energy spectrum.  For data runs during which the weather or the
detector performance caused a Cosmic Ray detection rate deviating only
slightly, i.e.\ less than 15\% from the expectation value, the
$\gamma$-ray detection rates and spectra were corrected accordingly.
Spectral results above an energy threshold of 500~GeV were derived
from the data of zenith angles smaller than 30$^\circ$ (39~hours of
data).  The determination of the diurnal integral flux estimates and
the search for variability within individual nights use all data.

\section{Results}

\subsection{Light curves}

Figure 1 shows the HEGRA and energy-dependent {\it RXTE} light curves
re-binned on 1 day and 5408 s ($\sim$ one orbit), respectively.  The
PCA light curves were accumulated in the energy ranges 2--4~keV and
10--20~keV; for an assumed spectrum with a typical
$\Gamma_{3-20~keV}=2.3$ (see below), their effective energies are
3~keV and 16 keV, respectively (not significantly dependent on the
slope).

After a period of very low activity at both TeV and X-rays, a strong
flare is apparent at all energies starting on day MJD 50989 and ending
on day MJD 50994. At TeV energies, the flare has a broad base, lasting
approximately six days, with a narrow ``core'' superposed, lasting two
days (MJD 50991--992), and a total max/min amplitude of a factor
$\sim$ 20. The X-rays track well the structure of the TeV flare,
although with lower amplitudes (factor 4 and 2 at hard and soft
X-rays, respectively). A correlation analysis using both the Discrete
Correlation Function and Modified Mean Deviation methods (Edelson \&
Krolik 1989; Hufnagel \& Bregman 1992) confirm that there are no lags
between the TeV and X-ray light curves, or between the soft and hard
X-rays, larger than one day.

To explore correlations on short timescales, we examined light curves
binned at 900 s in TeV and 300 s at X-rays (the best compromise
between time resolution and adequate signal-to-noise ratio in both
cases). Figure 2 shows the TeV and X-ray light curves for the day of
the peak activity, i.e., MJD 50991, when intra-hour variability at TeV
energies was detected. The TeV flux varied by a factor $\sim$ 2, with
the hypothesis of constant flux rejected at 99.4\% confidence level
according to the $\chi^2$ test.  The doubling timescale of the TeV
flux is well below 1~hour (approximately 20~min); to our knowledge,
this is the shortest flux variability timescale found for Mrk~501 so
far (e.g., Quinn et al. 1999), and comparable to Mrk 421 (Gaidos et
al. 1996). Unfortunately, as Figure 2 shows, gaps in the {\it RXTE}
sampling prevent us from commenting on sub-hour correlated variability
at X-rays. 

A very rapid X-ray flare, with an increase of the 2--10 keV flux by
60\% in $<$ 600 s, was recently detected from Mrk 501 with {\it RXTE}
in 1998 May (Catanese \& Sambruna 2000). This result, together with
our evidence for fast TeV variability, shows that Mrk 501 can vary on
the fastest timescales at both X-ray and TeV wavelengths as other TeV
sources (Mrk 421; Maraschi et al. 1999), and calls for future dense
X-ray/TeV monitorings, aimed at probing correlated variability on the
shortest accessible timescales.  

\subsection{Simultaneous TeV and X-ray spectra} 

Because of the sampling, we are able to derive truly simultaneous
X-ray and TeV spectra during the pre-flare and the flare states.  The
high-state spectra were accumulated during the days of maximum TeV
activity, MJD 50991--992, while the pre-flare spectra were accumulated
in the time interval MJD 50979--990.  Table 1 reports the results of
the spectral fitting of the simultaneous TeV and X-ray spectra, while
the data are shown in Figure 3.

The HEGRA spectrum during the flare state was fitted over the energy
range from 500~GeV to 20~TeV (above 10~TeV the evidence for emission
is only marginal) with a power law plus an exponential cutoff,
dN/dE=N$_0 \times$ (E/TeV)$^{-\Gamma} \times e^{(-E/E_0)}$, with
spectral parameters reported in Table 1 (with statistical
uncertainties). 
The parameters $E_0$ and $\Gamma$ are strongly correlated: within
systematic errors the pairs of parameters ($\Gamma=1.7$;
$E_0=2.8$~TeV) and ($\Gamma=2.2$; $E_0=6.6$~TeV) are also consistent
with the data.  Note that the spectral parameters we measure for the
1998 June outburst, i.e., a slope $\Gamma=1.9$ and cutoff energy
$E_0=4$ TeV, are very similar to those measured during the 1997
flaring phase (Aharonian et al.\ 1999a).  For the pre-flare phase, the
TeV-flux was too low to allow us to fit a power law model with an
exponential cutoff (Table 1).  A fit of a power law model to the ratio
of the flare and the pre-flare spectra gives $(d$N/$d$E)$_{\rm
flare}/(d$N/$d$E$)_{\rm pre-flare} \propto $E$^\beta$ with $\beta =
-0.17 \pm 0.19$, consistent within statistics with no spectral
evolution.

The PCA spectra were fitted with a single power law with Galactic
absorption, 1.73 $\times 10^{20}$ cm$^{-2}$ (Elvis, Lockman, \& Wilkes
1989). As can be seen from Table 1, this model provides an excellent
fit to the X-ray spectra up to 20 keV, with the photon index
flattening from $\Gamma_{3-20~keV}=2.21$ during the pre-flare state to
$\Gamma_{3-20~keV}=1.89$ during the flare. No spectral breaks are
required, i.e., there is no statistical improvement when a second
power law is added to the fit. However, we can not exclude the
presence of a spectral break at energies softer than sampled with the
PCA, $\sim$ 1--2 keV, as indeed detected with {\it BeppoSAX} (Pian et
al. 1998a).

The HEXTE data are fitted by a power law with a photon index
consistent with the extrapolation of the PCA slope in both high and
low states (Table 1). Indeed, fitting the PCA and HEXTE datasets
together, we find that a single power law with a slope similar to the
PCA slope describes well the 3--50 keV continuum during both the
pre-flare and flare epochs. Given the large uncertainties of the HEXTE
data, however, we can not rule out the presence of spectral breaks at
energies $\gtrsim 10-20$ keV, as indeed detected by {\it BeppoSAX}
(Pian et al. 1998a,b).

\subsection{X-ray and TeV spectral variability} 

We accumulated time-resolved PCA spectra for each data point of the
X-ray light curves in Figure 1, and fitted them over the energy range
3--20 keV with a single power law plus Galactic absorption. The
results of the fitting are reported in Table 2 (columns 3--5),
together with the date of the spectrum (column 1) and its net exposure
(column 2). The time progression of the PCA slope is plotted in Figure
1, intermediate panel.  Large variability is readily apparent, with
the photon index flattening from $\Gamma_{3-20~keV} \sim 2.3$ to
$\Gamma_{3-20~keV} \sim 1.8$ with increasing flux.  There is an
indication that the X-ray continuum steepens during the decay stage of
the flare.

The X-ray spectral variations follow a well-defined pattern with the
intensity. This is illustrated in Figure 4, where the 3--20 keV photon
index is plotted versus the 2--10 keV flux.  The dotted lines mark the
time progression of the slope during the flaring activity, and clearly
show a ``clock-wise'' loop. This is similar to what was observed in
other HBLs (PKS 2005--489, Perlman et al. 1999; PKS 2155--304, Sembay
et al. 1992, Sambruna 1999; Mrk 421, Takahashi et al. 1996) and can be
interpreted in terms of cooling of the synchrotron-emitting electrons
in the jet (Kirk, Riegler, \& Mastichiadis 1998).

The HEXTE spectrum accumulated at the beginning of the flare (see
\S~2) is fitted by a power law with slope $\Gamma_{20-50~keV}=2.19 \pm
0.59$ and 20--50 keV flux F$_{20-50~keV}=(5.3 \pm 1.6) \times
10^{-11}$ erg cm$^{-2}$ s$^{-1}$. During the peak and decreasing
flare, $\Gamma_{20-50~keV}=1.86 \pm 0.28$ and F$_{20-50~keV}=(1.1 \pm
0.2) \times 10^{-10}$ erg cm$^{-2}$ s$^{-1}$. Comparing to the
pre-flare flux from Table 1, the source brightened by a factor $\sim$
6 during the TeV flare in the HEXTE band, with an indication of a
hardening of the 20--50 keV continuum.

At TeV energies, given the limited signal-to-noise ratio in the
pre-flare state, we investigated spectral variations by constructing
hardness ratios. These are defined as the ratios of the flux in
2--9.7~TeV to the flux in 0.8--2~TeV (the lower bound is chosen to
assure negligible systematic errors due to threshold effects and 2~TeV
approximately equals the median energy of photons with energies above
0.8~TeV).  The TeV hardness ratios are plotted versus the observation
date in Figure 1, bottom panel, together with 1$\sigma$ uncertainties.
It is apparent that, within statistical uncertainties, the hardness
ratios in the pre-flare state (MJD 50979--990) and flare state (MJD
50991--992) are very similar, despite that the absolute fluxes differ
by one order of magnitude.  Intriguingly, the spectrum seems to soften
substantially during the decay stage, although the limited statistical
significance of about 2$\sigma$ prevents us from drawing firmer
conclusions.

\section{Discussion}

Since the typical flux variability timescale of Mrk 501 in TeV
$\gamma$-rays and X-rays can be much less than one day, it is
important to have truly simultaneous observations in both bands. It is
also important to have reasonably continuous sampling on timescales of
at least one day in order to have an accurate picture of the dynamics
of the source.  For this reason, we conducted a 15-day TeV/X-ray
monitoring with diurnal {\it RXTE} observations exactly in the HEGRA
visibility windows.  After 10 days of quiescence, the source exhibited
a strong flare at both TeV and X-rays lasting six days, with a flux
exceeding the pre-flare level by a factor of $\sim $~20 at TeV
energies during a 2-day maximum, and with lower amplitudes (factor
2--4) at X-rays.  We also report the first detection of TeV flux
variability on sub-hour timescales in Mrk 501 (\S~3.1).

By chance, our multiwavelength campaign in 1998 June coincided with
the only high TeV activity of the source during that year. Luckily, we
were able to follow the evolution of the TeV flare not only during the
pre-flare and flare stage but also during the decay stage. 
The TeV spectrum during the flare is similar to the spectra observed
in 1997, suggesting that the flaring episode we witnessed in 1998 June
was a scaled-down version of the longer-lasting 1997 flare.  This
conclusion is bolstered by the strong spectral variations we observe
in the X-rays. Our {\it RXTE} observations show that the X-ray
continuum in 3--20 keV flattened by $\Delta\Gamma_{3-20~keV} \sim 0.5$
from the beginning of the campaign ($\Gamma_{3-20~keV}=2.3$) to the
flare maximum ($\Gamma_{3-20~keV}=1.8$).  Interestingly, at the peak
of the TeV flare the X-ray slope was similar to the 2--10 keV slope
measured in 1997 April, May, and July with {\it BeppoSAX} and {\it
RXTE} (Pian et al. 1998a; Lamer \& Wagner 1999; Krawczynski et
al. 1999).  This implies a similar shift of the synchrotron peak
frequency at higher energies, $\gtrsim 50$ keV (Figure 3).  While in
April 1997 the X-ray continuum flattened by 0.4 within approximately
two weeks, we see here a comparable flattening within only $\sim$ 2--3
days. Note that large changes of the position of the synchrotron peak
are relatively rare. Besides Mrk 501, they were observed to-date only
in two HBLs, 1ES 2344+514 (Giommi, Padovani, \& Perlman 1999) and 1ES
1426+428 (Ghisellini, Tagliaferri, \& Giommi 1999), but not in Mrk
421, PKS 2155--304, or any other BL Lac. Our observations provide the
first evidence that in Mrk 501 the synchrotron peak may change on
relatively short timescales ($\sim$ a few days). 

Several models have been suggested to explain the TeV radiation from
blazars. A popular scenario are the leptonic models, where TeV
$\gamma$-rays are produced via inverse Compton scattering of directly
accelerated electrons on external and/or internal photons (e.g.,
Sikora 1997).  For Mrk 501, an object without strong broad line
emission, the synchrotron self-Compton (SSC) model is almost commonly
accepted as the most probable explanation for the observed
X-ray/TeV-$\gamma$-ray emission (e.g., Tavecchio et al. 1998; Kataoka
et al. 1999).  Presently, the SSC model is the only model (at least in
its simplified, ``one-zone'' version) which has been developed to a
level which allows conclusive predictions which can be compared with
experimental results.  In particular, the SSC scenario is able to give
satisfactory fits to both the X-ray and the TeV spectra (Pian et
al. 1998a; Hillas 1999; Krawczynski et al. 1999). We used the code
developed by Coppi (1992) to fit our simultaneous SEDs in Figure 3,
assuming emission from a one-zone, homogeneous region and
incorporating Klein-Nishina effects. The key parameters used in this
model are the Doppler factor $\delta_{\rm j}$ of the relativistic
plasma, the radius $R$ of the emission region, the magnetic field $B$,
and the electrons' maximum energy $E_{max}$.

The results of the fits are shown in Figure 3 as solid lines, and the
parameters' values are reported in the caption.  As discussed further
below, the models were computed without correcting for the
extragalactic extinction of TeV photons due to $\gamma$/$\gamma$ pair
production with the photons of the Diffuse Extragalactic Background
Radiation (e.g.\ Gould \& Schr\'{e}der 1966).  In the lower panel, we
plot the ratio of the data and best-fit model between the high-state
and pre-flare. The latter plot emphasizes that, while the TeV spectra
of both states are quite similar, the X-ray spectra of the pre-flare
and the flare state are significantly different.  In SSC models the
hardening of the X-ray spectrum during the flare can be attributed to
a shift of the peak frequency $\nu_{\rm s}$ of the synchrotron
radiation, $\nu_{\rm s}\propto~B \times E^2_{\rm max}$.  Assuming an
increase of both the magnetic field and the maximum energy during the
flare, the dramatic changes of the X-ray spectrum are readily
explained (see Figure 3).  While the increase of magnetic field does
not affect the $\gamma$-ray spectrum, the increase of $E_{\rm max}$
does make the inverse Compton (IC) spectrum harder.  However, since
the $\gamma$-rays are produced in the Klein-Nishina regime, this
effect is less pronounced in IC than in the synchrotron radiation
component.  The fits shown in Figure 3 correspond to the following
model parameters: $B$=0.03~G, $E_{\rm max}=2$~TeV (exponential cutoff
energy), $R=4 \times 10^{16}$\,cm for pre-flare state, and $B$=0.1~G,
$E_{\rm max}=20$~TeV, $R=2.7 \times 10^{15}$\,cm for the flare state.
For both cases a Doppler factor of $\delta_{\rm j}=25$ is assumed.
Note that the latter value of the Doppler factor implies that internal
absorption of the TeV $\gamma$-rays by lower frequency photons can be
completely neglected (e.g.\ Celotti, Fabian, \& Rees 1998).
Furthermore, the chosen Doppler factor and radius of the emitting
volume in the flaring state imply time variability down to
$t=R/(c~\delta_{\rm j}) =$1~hour which agrees with the observed flux
variability following from Figure 2.  For the flare state with good
statistics up to $\sim$ 10~TeV, the model over-predicts the TeV flux
above $\sim$ 5~TeV, in particular by a factor of $\sim$ 2.5 at 10~TeV.
This discrepancy should not be overemphasized, but could well be the
result of intergalactic extinction due to $\gamma$/$\gamma$ pair
production.

In summary, we have performed a 2-week monitoring campaign of the HBL
Mrk 501 in 1998 June with HEGRA and {\it RXTE}, with a sampling
designed to probe TeV/X-ray correlation on timescales of several
hours. We detected a strong flare at both wavelengths, rising from a
period of very low activity, well correlated at TeV and X-rays on time
scales of $\lesssim$ 1 day, accompanied by large
($\Delta\Gamma_{3-20~keV} \sim 0.5$) spectral variability at
X-rays. Our results support an interpretation in terms of a canonical
synchrotron-self Compton scenario. Future campaigns with a more
intensive sampling designed to probe correlation on shorter time
scales at both X-ray and TeV energies are needed to set more stringent
constraints on the radiative processes which play an important role in
the evolution of the flare.

\acknowledgements 

RMS acknowledges support from NASA contract NAS--38252 and NASA grant
NAG--7121. RR acknowledges support by NASA contract NAS5-30720. We are
grateful to the {\it RXTE} team, especially Evan Smith, for making
these observations possible, to the {\it RXTE} GOF for constant
support with the data analysis, and to Joe Pesce for a careful reading
of the manuscript. HEGRA is supported by the German ministry for
Research and technology BMBF and the Spanish Research Council CICYT.
We thank the Instituto de Astrophys\`{i}ca de Canarias for supplying
excellent working conditions at La Palma.  HEGRA gratefully
acknowledges the technical support staff of the Heidelberg, Kiel,
Munich, and Yerevan Institutes.

\newpage

\noindent{\bf Figure Captions}

\begin{itemize}
 
\item\noindent Figure 1: Multiwavelength light curves of Mrk 501 in
1998 June, as measured with HEGRA and {\it RXTE}, binned at 1 day and
5408 s (one orbit), respectively (top panel). The HEGRA flux units are
$10^{-12}$ ph cm$^{-2}$ s$^{-1}$, the {\it RXTE} data are in c
s$^{-1}$. The HEGRA light curve was arbitrarily shifted by +1.5 in
logarithmic units for clarity of presentation.  A strong flare is
detected at both TeV and X-rays, with increasing amplitude for
increasing energy. The flare was accompanied by large spectral
variations at X-rays (middle panel), with flatter slope with
increasing flux. Within the statistical errors, the TeV spectrum was
rather hard during the whole pre-flare and flare phases, as shown by
the TeV hardness ratios in the bottom panel (upper limits are on
1$\sigma$ confidence limit to facilitate the comparison with the error
bars of the flux estimates).  There is an indication of spectral
softening during the decay stage of the flare.

\item\noindent Figure 2: TeV and X-ray light curves (binned at 900 s 
and 300 s, respectively) of Mrk 501 during the day of maximum TeV
activity in 1998 June. Significant variability of a factor $\sim$2 on
$\sim$ 20 min timescale is detected at TeV energies. Unfortunately,
gaps are present in the {\it RXTE} monitoring and we can not comment
on correlated X-ray variability on these short timescales.

\item\noindent Figure 3: Spectral energy distributions of Mrk 501 in
1998 June during the peak of the TeV/X-ray flare (filled dots) and
during the pre-flare state (open dots). Only the PCA data are plotted
for clarity (Table 1). The solid lines are fits to the spectra with an
homogeneous SSC model (Coppi 1992), with the following fitted
parameters: $B$=0.03~G, $E_{\rm max}=2$~TeV, $R=4 \times 10^{16}$ cm
for the pre-flare state; $B$=0.1~G, $E_{\rm max}=20$~TeV, $R=2.7
\times 10^{15}$ cm for the high state.  The bottom panel shows the
ratios of the model spectra and data for the flare and pre-flare
states.

\item\noindent Figure 4: Plot of the X-ray 3--20 keV slope versus the
observed 2--10 keV flux, from fits to the time-resolved PCA spectra
(Table 2). The trend of flattening slope with increasing flux is
apparent. The dotted lines mark the time progression of the slope,
which appears to follow a ``clock-wise'' pattern during the
flare. This behavior is consistent with the X-ray flare being due to
electron cooling (Kirk et al. 1998). 

\end{itemize}

\newpage 

\scriptsize

\oddsidemargin-0.85in
\textheight10.2in
\textwidth7.7in
\topmargin-0.8in
\footheight0in                                    
\footskip0in                                      
~~~~ \\

\begin{center}
\begin{tabular}{lcccccc}
\multicolumn{7}{l}{{\bf Table 1: Simultaneous average TeV and X-ray spectra}} \\
\multicolumn{7}{l}{   } \\ \hline
& & & & & & \\
State$^{~a}$ && N$_0^{~b}$ & $\Gamma$ & E$_0$ &  F$^{~c}$ & $\chi^2_r$/dofs \\
   & & & & (TeV) & (10$^{-11}$ erg cm$^{-2}$ s$^{-1}$) &  \\
& & & & & & \\ \hline
& & & & & & \\
\multicolumn{7}{l}{{\bf A) TeV$^{~d}$ }} \\
& & & & & & \\
Flare && 7.9 $\pm$ 1.0 & 1.92 $\pm$ 0.3 & 4.0$^{+1.45}_{-0.90}$ &$\cdots$  & 0.54/13 \\
Pre-flare && 0.5 $\pm$ 0.1 & 2.31 $\pm$ 0.20 & $\cdots$  &$\cdots$ & 1.4/9 \\ 
& & & & & & \\
\multicolumn{7}{l}{{\bf B) X-ray$^{~e}$ }} \\
& & & & & & \\
Flare PCA && $\cdots$ & 1.89 $\pm 0.02$ &$\cdots$ & 18.5 $\pm$ 0.9 & 0.75/42 \\
Flare HEXTE && $\cdots$ & 2.19 $\pm$ 0.29 & $\cdots$ & 7.5 $\pm$ 1.1 & 1.04/69 \\
Pre-flare PCA && $\cdots$ & 2.21 $\pm$ 0.02 &$\cdots$ & 0.7 $\pm$ 0.1 & 0.85/41 \\ 
Pre-flare HEXTE && $\cdots$ & 2.30 $\pm$ 0.45 & $\cdots$ & 1.8 $\pm$ 0.4 & 0.87/69 \\
& & & & & & \\ \hline

\end{tabular}
\end{center}

\indent\indent\indent{\bf Notes:} \\
\indent\indent\indent a=High state corresponds to the time interval MJD 50991--992. Low state corresponds to MJD 50979--987; \\
\indent\indent\indent b=Normalization of the power law, in 10$^{-11}$ ph cm$^{-2}$s$^{-1}$TeV$^{-1}$ for the HEGRA data; \\
\indent\indent\indent c=Observed flux in 2--10 keV (PCA) and 20--50 keV (HEXTE); \\
\indent\indent\indent d=Fits with a power law plus
exponentional cutoff: dN/dE=N$_0 \times$ (E/TeV)$^{-\Gamma} \times
e^{(-E/E_0)}$. Errors on \\
\indent\indent\indent\indent parameters are statistical; \\
\indent\indent\indent e=Fits with a single power law plus Galactic
absorption, N$_H=1.73 \times 10^{20}$ cm$^{-2}$ (Elvis et al. 1989). \\

\newpage 

\oddsidemargin-0.85in
\textheight10.2in
\textwidth7.7in
\topmargin-0.8in
\footheight0in                                    
\footskip0in                                      
~~~~ \\

\begin{center}
\begin{tabular}{lrccc}
\multicolumn{5}{l}{{\bf Table 2: X-ray spectral variability$^{~a}$}} \\
\multicolumn{5}{l}{   } \\ \hline
& & & & \\
Start Date & Net Exp. & $\Gamma_{3-20~keV}$ & $\chi^2_r$ & F$_{2-10~keV}$ \\
(MJD-50000) & (s) &   & (for 42 dofs)  & (10$^{-11}$ erg cm$^{-2}$ s$^{-1}$) \\& & & & \\ \hline 
& & & & \\
978.9 & 3168 & 2.29 $\pm$ 0.04 & 0.55 & 5.91 \\
979.9 & 3312 & 2.27 $\pm$ 0.04 & 0.72 & 6.05 \\
980.9 & 6304 & 2.31 $\pm$ 0.03 & 0.57 & 6.04 \\
981.9 & 6320 & 2.17 $\pm$ 0.03 & 0.84 & 7.13 \\
982.9 & 3488 & 2.22 $\pm$ 0.04 & 0.66 & 6.29 \\
983.0 & 4144 & 2.23 $\pm$ 0.04 & 0.46 & 6.41 \\
983.9 & 6192 & 2.21 $\pm$ 0.03 & 0.73 & 8.41 \\
984.0 & 1328 & 2.19 $\pm$ 0.06 & 0.70 & 6.58 \\
984.9 & 5040 & 2.16 $\pm$ 0.03 & 0.70 & 6.08 \\
985.1 & 464  & 2.26 $\pm$ 0.10 & 0.68 & 6.02 \\
985.9 & 3024 & 2.19 $\pm$ 0.04 & 0.71 & 6.47 \\
986.0 & 352  & 2.07 $\pm$ 0.11 & 0.77 & 6.43 \\
986.1 & 528  & 2.25 $\pm$ 0.09 & 0.78 & 6.45 \\
986.9 & 384  & 2.21 $\pm$ 0.10 & 0.68 & 6.71 \\
987.1 & 480  & 2.36 $\pm$ 0.10 & 0.78 & 6.76 \\
987.9 & 1536 & 2.28 $\pm$ 0.06 & 1.04 & 7.05 \\
988.0 & 592  & 2.20 $\pm$ 0.08 & 0.71 & 7.29 \\
988.9 & 1152 & 2.06 $\pm$ 0.04 & 0.71 & 10.8 \\
989.0 & 912  & 2.06 $\pm$ 0.04 & 0.60 & 11.4 \\
989.9 & 1296 & 1.96 $\pm$ 0.03 & 0.65 & 12.1 \\
990.0 & 208  & 2.03 $\pm$ 0.08 & 0.52 & 11.3 \\
990.0 & 512  & 2.06 $\pm$ 0.06 & 0.74 & 11.3 \\
990.9 & 1392 & 1.89 $\pm$ 0.02 & 1.46$^{~b}$ & 16.9 \\
991.0 & 656  & 1.86 $\pm$ 0.03 & 0.99 & 18.1 \\
991.9 & 432  & 1.91 $\pm$ 0.03 & 0.52 & 20.5 \\
992.0 & 512  & 1.93 $\pm$ 0.04 & 0.65 & 20.0 \\
992.9 & 528  & 2.01 $\pm$ 0.04 & 0.47 & 15.4 \\
993.0 & 736  & 2.08 $\pm$ 0.04 & 0.82 & 15.4 \\
& & & & \\ \hline 

\end{tabular}
\end{center}

\indent\indent\indent{\bf Notes:} \\
\indent\indent\indent a=Fits to the PCA data in 3--20 keV with a single
power law plus Galactic N$_H$ (1.73 $\times 10^{20}$ cm$^{-2}$). \\
\indent\indent\indent\indent Errors are 
90\% confidence for one parameter of interest
($\Delta\chi^2$=2.7). \\
\indent\indent\indent b=High $\chi^2$ is due to instrumental absorption 
features in the residuals around 4.8 keV (Xenon edge) \\
\indent\indent\indent\indent and 8.5 keV (unknown origin). 

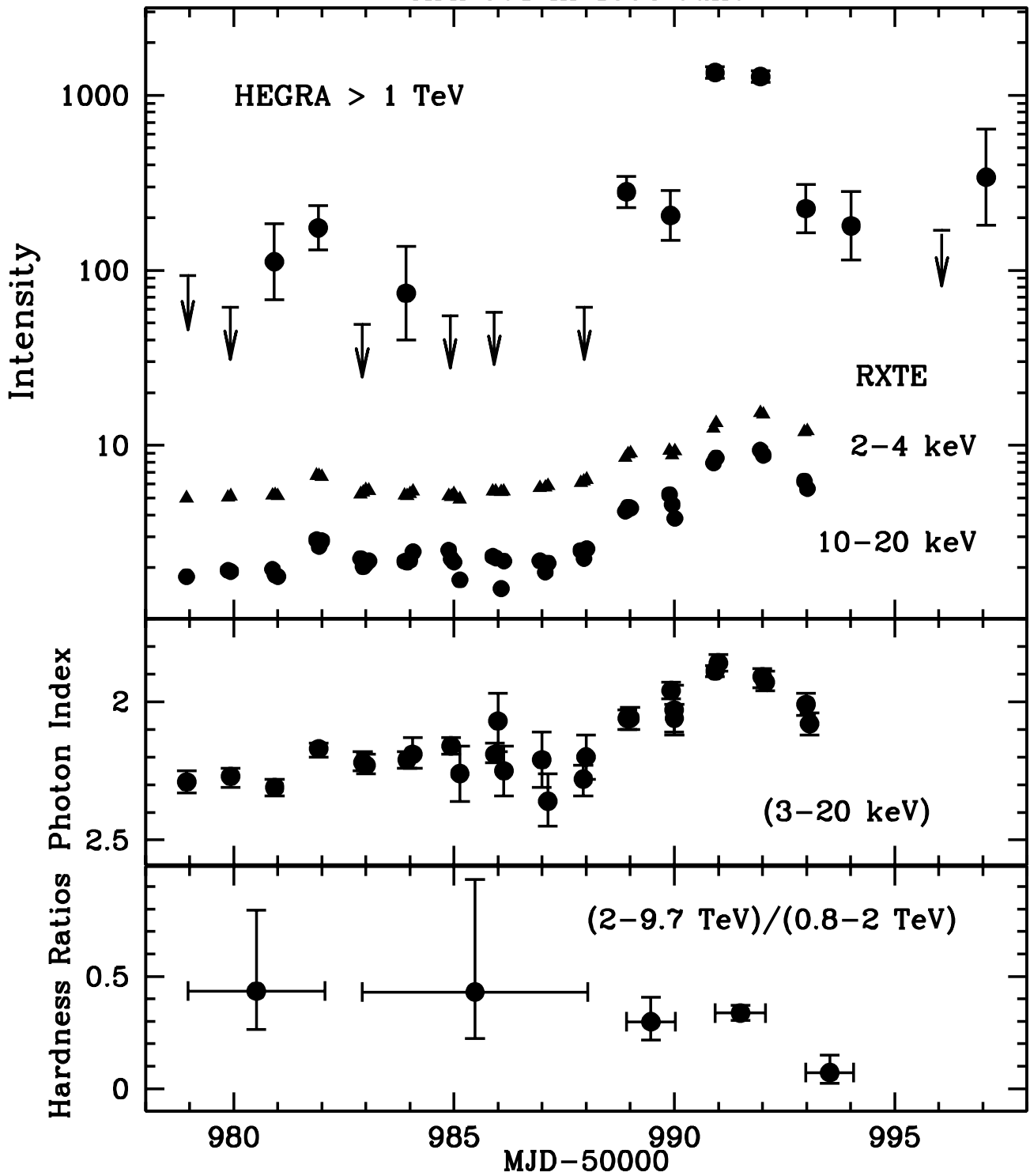
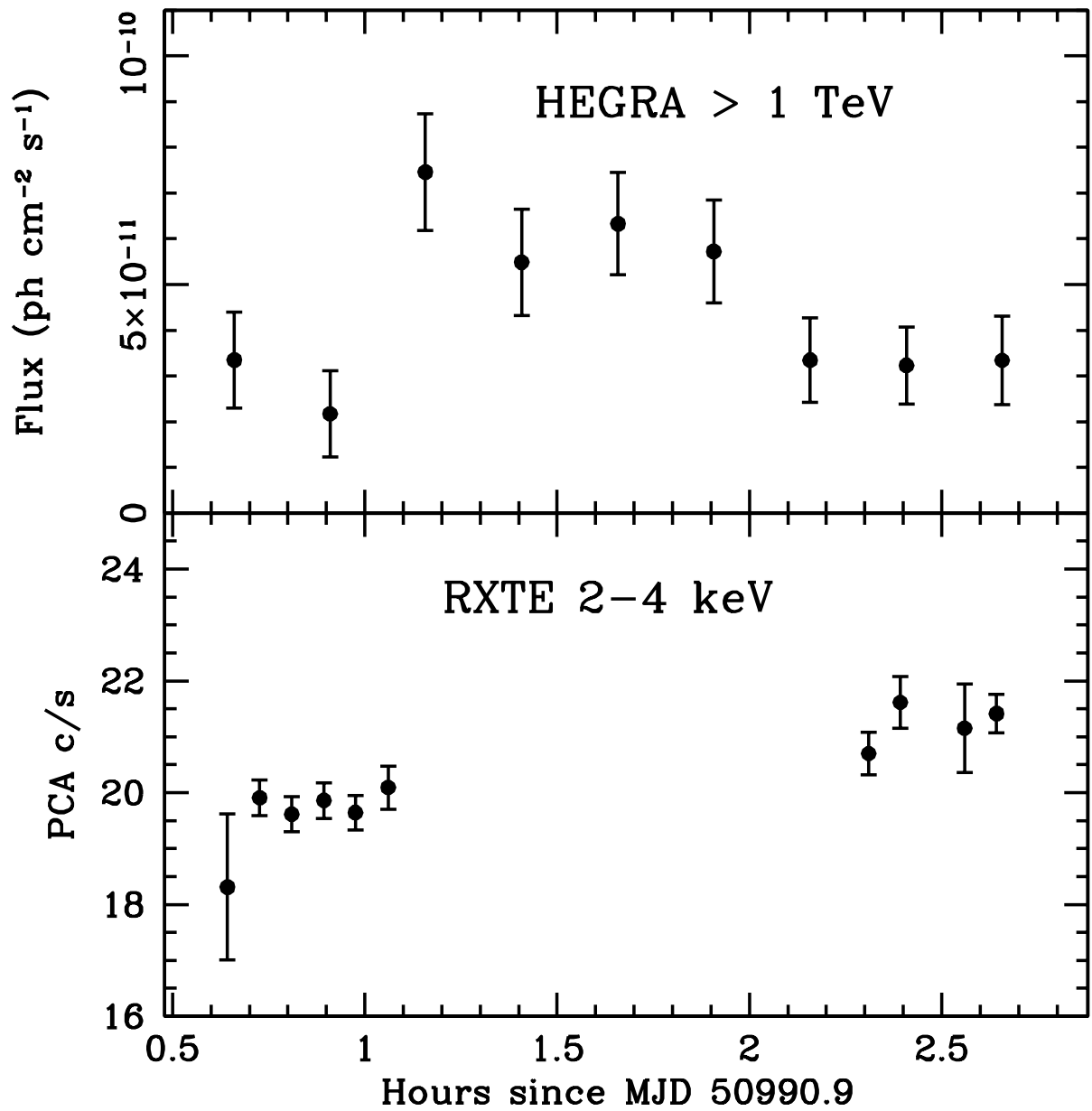
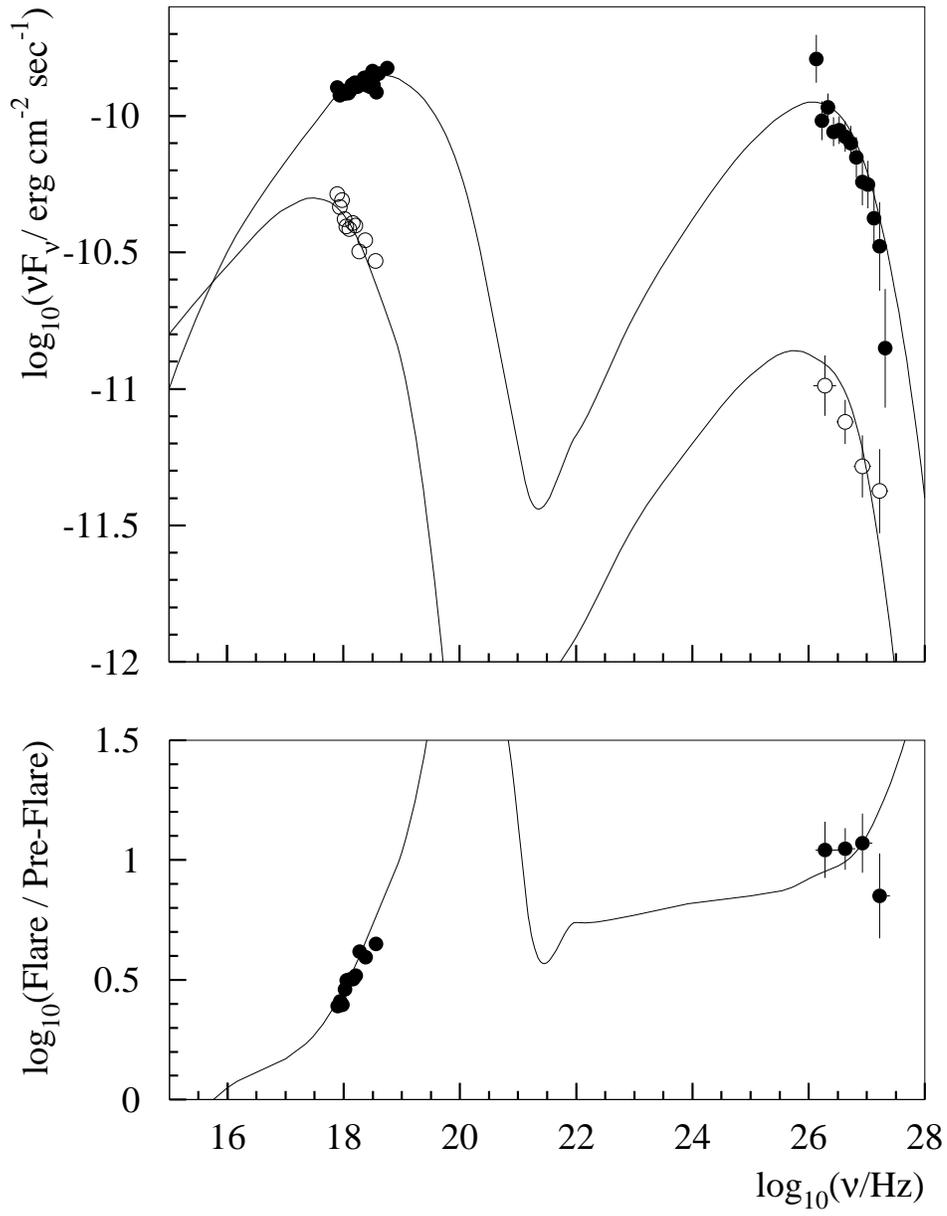
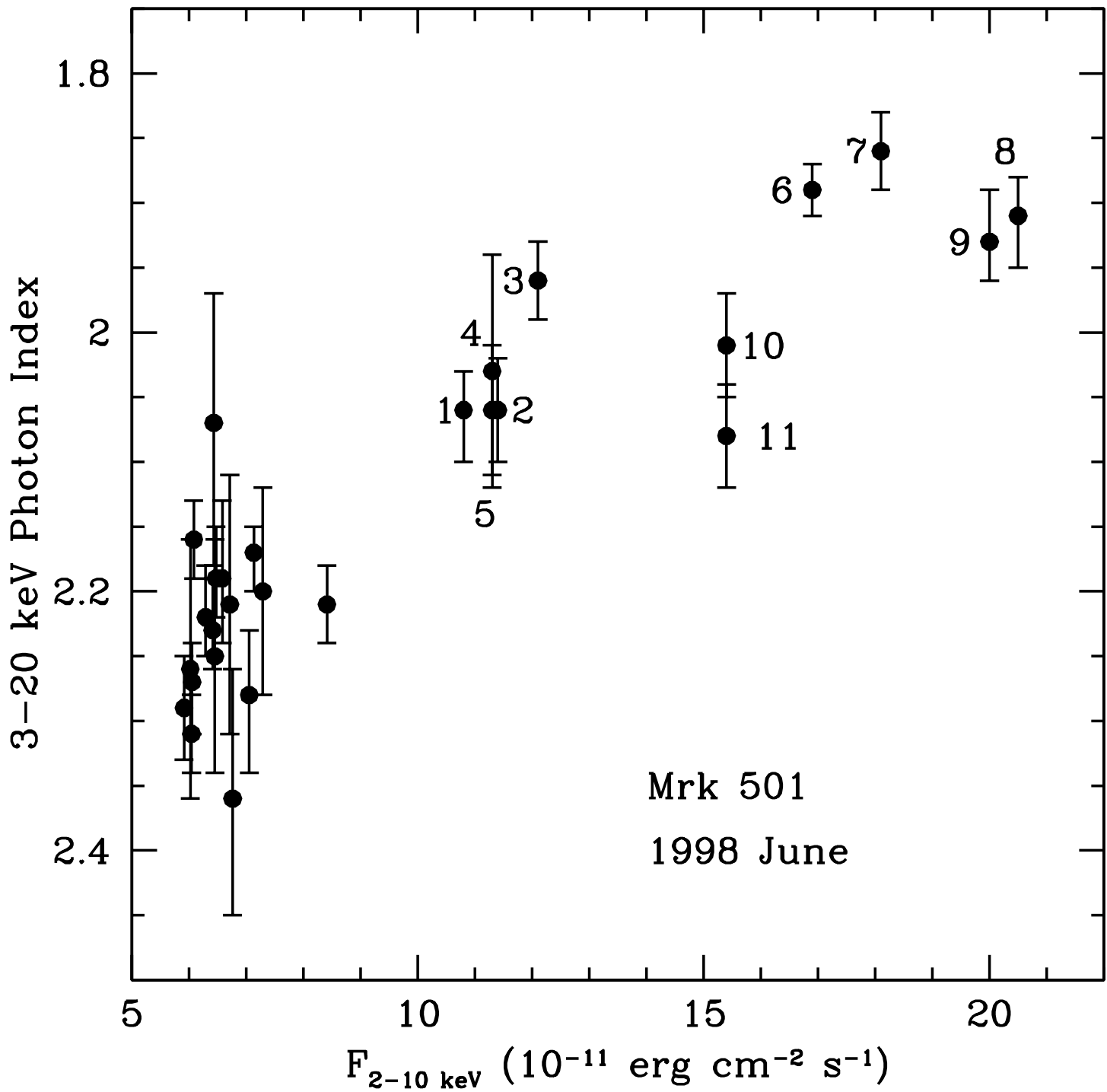
\end{document}